\documentclass[a4paper,10pt]{article}
\usepackage[utf8]{inputenc}

\usepackage[spanish,english]{babel}
\usepackage{graphicx}
\usepackage{amssymb}
\usepackage{amsmath}
\usepackage{color}
\usepackage{hyperref}
\usepackage{latexsym,dsfont,mathtools}
\usepackage{theorem,titletoc,titlesec}
\usepackage{enumitem}
\usepackage{tikz}
\usepackage{nccmath}
\usepackage{subcaption}
\usepackage{ulem}
\usepackage{pgfplots}
\usepackage[ruled]{algorithm}
\usepackage{algorithmic}
\usepackage{comment}

\DeclareMathAlphabet{\mathpzc}{OT1}{pzc}{m}{it}

\usetikzlibrary{matrix}

\pgfplotsset{width=7cm,compat=1.16}

\newtheorem{theorem}{Theorem}[section]
\newtheorem{corollary}{Corollary}[section]
\newtheorem{proposition}{Proposition}[section]
\newtheorem{lemma}{Lemma}[section]
\newtheorem{example}{Example}[section]
\newtheorem{remark}{Remark}[section]
\newtheorem{definition}{Definition}[section]

\newenvironment{proof}[1][Proof.]{\vspace{0.5em}\textbf{#1} }{\
\hfill$\mathcal{QED}$}

    {\end{pmatrix}\end{medsize}}%

\newcommand{\Z}{\mathbb{Z}}
\newcommand{\zero}{{\mathbf{0}}}

\newcommand{\C}{{\cal C}}

\newcommand{\zps}{\mathbb{Z}_{p^s}}
\newcommand{\cS}{{\cal S}}

\begin{document}

\title{Computing efficiently a parity-check matrix for $\Z_{p^s}$-additive codes.\thanks{This work has been partially supported by the Spanish MINECO under grants  PID2019-104664GB-I00,   PID2022-137924NB-I00, and RED2022-134306-T
(AEI / 10.13039/501100011033), by the Catalan AGAUR under grant 2021 SGR 00643, and by the Portuguese Foundation for Science and Technology (FCT-Fundação para a Ciência e a Tecnologia), through CIDMA - Center for Research and Development in Mathematics and Applications, within project UIDB/04106/2020.}
}

\author{Cristina Fern\'andez-C\'ordoba \and Adri{\'a}n Torres \and Carlos Vela \and Merc\`e Villanueva}


\maketitle
\begin{abstract}
The $\Z_{p^s}$-additive codes of length $n$ are subgroups of $\Z_{p^s}^n$, and can be seen as a generalization of linear
codes over $\Z_2$, $\Z_4$, or more general over $\Z_{2^s}$. In this paper, we show two methods for computing a parity-check matrix of a 
$\Z_{p^s}$-additive code from a generator matrix of the code in standard form. We also compare the performance of our results implemented in Magma with the current available function in Magma for codes over finite rings in general. A time complexity analysis is also shown.  
\end{abstract}

\section{Introduction}

Let $\Z_{p^s}$ be the ring of integers modulo $p^s$ with $p$ prime and $s\geq1$. The set of
$n$-tuples over $\Z_{p^s}$ is denoted by $\Z_{p^s}^n$. In this paper,
the elements of $\Z^n_{p^s}$ will also be called vectors. 
The order of a vector $u$ over $\Z_{p^s}$, denoted by $o(u)$, is the smallest positive integer $m$ such that $m u =\zero$.

A code over $\Z_p$ of length $n$ is a nonempty subset of $\Z_p^n$,
and it is linear if it is a subspace of $\Z_{p}^n$. Similarly, a nonempty
subset of $\Z_{p^s}^n$ is a $\Z_{p^s}$-additive if it is a subgroup of $\Z_{p^s}^n$. 
Note that, when $p=2$ and $s=1$, a $\Z_{p^s}$-additive code is a binary linear code and, when $p=2$ and $s=2$ ,
it is a quaternary linear code or a linear code over $\Z_4$. $\Z_{p^s}$-additive codes were first defined as $p$-adic codes in \cite{Calderbank}. Let $\C$ be a $\Z_{p^s}$-additive code of length $n$.
Since $\C$ is a subgroup of $\Z_{p^s}^n$, it is isomorphic to an abelian group $\Z_{p^s}^{t_1}\times\Z_{p^{s-1}}^{t_2}\times \dots\times\Z_p^{t_s}$, and we say that $\C$ is of type $(p^s)^{t_1}(p^{s-1})^{t_2}\dots p^{t_s}$ or briefly $(n;t_1,\dots,t_s)$. It is clear that a $\Z_{p^s}$-additive code of type $(n;t_1,\dots,t_s)$ has $p^{st_1+(s-1)t_2+\cdots+t_s}$ codewords. These codes have been considered and studied for example in \cite{Carlet,Gupta,Krotov2007,Shi2018,Shi2021}.  

In general, $\Z_{p^s}$-additive codes are not free as submodules of $\Z_{p^s}^n$ (they are free only when $t_2=\cdots=t_s=0$), which means that they usually do not have a basis, that is, a generating set consisting of linearly independent vectors. However, there exists a set of codewords $S=\{\mathbf{u}_i^{(j)} \mid 1\leq j\leq s, 1\leq i \leq t_j\}\subseteq\C$, where $o(\mathbf{u}_i^{(j)})=p^{s-j+1}$ for all $i$ and $j$, satisfying that any codeword in $\C$ can be expressed uniquely in the form
\begin{equation}
    \sum_{j=1}^s\sum_{i=1}^{t_j}\lambda_i^{(j)}\mathbf{u}_i^{(j)},
\end{equation}
for $\lambda_i^{(j)}\in\Z_{p^{s-j+1}}$. The matrix whose rows are the codewords in $S$ is a generator matrix of $\C$ having a minimum number of rows, that is, $t_1+\cdots+t_s$ rows.

The inner product of $\mathbf{u}=(u_1,\dots,u_n)$ and $\mathbf{v}=(v_1,\dots,v_n)$ in $\Z_{p^s}^n$ is defined as $\mathbf{u}\cdot \mathbf{v}=\sum_{i=1}^n u_iv_i\in\Z_{p^s}$. Then, if $\C$ is a $\Z_{p^s}$-additive code of length $n$, its dual code is
$$
\C^\perp=\{\mathbf{v}\in\Z_{p^s}^n\mid \mathbf{u}\cdot \mathbf{v}=0 \textnormal{ for all }\mathbf{u}\in \C\}.
$$
In \cite{Calderbank}, it is proved that if $\C$ is a $\Z_{p^s}$-additive code of type $(n;t_1,\dots,t_s)$, then $\C^\perp$ is a $\Z_{p^s}$-additive code of type $(n;n-t,t_s,t_{s-1},\dots,t_2)$, where $t=\sum_{i=1}^s t_i$.

Let $\C$ be a $\Z_{p^s}$-additive code with generator matrix $G$. A matrix $H$ is a parity-check matrix of $\C$ if it is a generator matrix of its dual code $\mathcal{C}^\perp$. In this sense, the code $\mathcal{C}$ can be generated from $H$ by computing all the orthogonal vectors to it, that is,
$$
\mathcal{C}=\{\mathbf{v}\in\Z_{p^s}^n\mid H\mathbf{v}^T=\mathbf{0}\}.
$$
A parity-check matrix $H$ holds that $GH^T=(\zero)$, which is a crucial property that plays the main role in syndrome decoding. It can be used to correct errors but also to correct erasures, since it provides a linear system of equations that can be solved to recover the sent information.

Two codes $C_1$ and $C_2$ over $\Z_p$ of length $n$ are said to be monomially equivalent (or just equivalent) provided there is a monomial matrix $M$ such that $C_2=\{ \textbf{c} M \mid \textbf{c} \in C_1 \}$. Recall that a monomial matrix is a square matrix with exactly one nonzero entry in
each row and column. They are said to be permutation equivalent if there is a permutation matrix $P$ such that $C_2=\{ \textbf{c}P \mid \textbf{c} \in C_1\}$. Recall that a permutation matrix is a square matrix with exactly one 1 in each row and column and $0$s elsewhere. 
Let $\cS_n$ be the symmetric group of permutations on the set $\{1,\dots,n\}$. A permutation matrix represents a permutation of coordinates, so we can also say that they are permutation equivalent if there is a permutation of coordinates $\pi \in \cS_n$
such that $C_2=\{ \pi(\textbf{c}) \mid \textbf{c} \in C_1 \}$.
Similarly, two $\Z_{p^s}$-additive codes, $\C_1$ and $\C_2$, of length $n$ are said to be permutation equivalent if they differ
only by a permutation of coordinates, that is, if there is a permutation of coordinates $\pi \in \cS_n$
such that $\C_2=\{ \pi(\textbf{c}) \mid \textbf{c} \in \C_1 \}$.

Let $Id_k$ be the identity matrix of size $k\times k$. In \cite{Sole}, it is shown that any quaternary linear code
code of type $(n;t_1,t_2)$ is permutation equivalent to a quaternary linear code with a generator
matrix of the form 
\begin{equation}\label{eq:GSF-4}
    G= \left ( \begin{array}{ccc}
 Id_{t_1} & R & S  \\
 \zero & 2Id_{t_2} & 2T    
 \end{array} \right ), 
\end{equation}
\noindent where $R$ and $T$ are matrices over $\Z_4$ with all
entries in $\{0,1\}\subset \Z_4$, and $S$ is a matrix over $\Z_4$. In this case, we say that $G$ is in standard form. In the same paper, it is shown that if the generator matrix $G$ of a quaternary linear code $\C$ is as in (\ref{eq:GSF-4}), then a parity-check matrix of $\C$ can be computed as follows:   
\begin{equation}\label{eq:HSF-4} 
H=\left (
\begin{array}{ccc}
-(S+RT)^{T} & T^{T}     & Id_{n-t_1-t_2} \\
2R^{T}  & 2Id_{t_2}& \zero\\ 
\end{array} \right ).
\end{equation}

In \cite{Calderbank}, a generator matrix in standard form for $\Z_{p^s}$-additive codes is given as a generalization of matrix (\ref{eq:GSF-4}) for quaternary linear codes. Specifically, if $\C$ is a $\Z_{p^s}$-additive code of type $(n;t_1,\dots,t_s)$, then $\C$ is permutation equivalent to a $\Z_{p^s}$-additive code $\C'$  generated by a generator matrix in standard form, that is, of the form
\begin{equation}\label{Gmatrix}
G=\left(\begin{array}{cccccc}
Id_{t_1} & A_{1,2} & A_{1,3} &\cdots & A_{1,s} & A_{1,s+1} \\
\zero & pId_{t_2} & pA_{2,3} & \cdots & pA_{2,s} & pA_{2,s+1}\\
\zero & \zero & p^2Id_{t_3} & \cdots & p^2A_{3,s} & p^2A_{3,s+1}\\
\vdots & \vdots & \vdots & \ddots & \vdots & \vdots\\
\zero & \zero & \zero & \cdots & p^{s-1}Id_{t_s} & p^{s-1}A_{s,s+1}
\end{array}\right),
\end{equation}
where $A_{i,j}$ are matrices over $\Z_{p^s}$ for all $1\leq j \leq s$ and $1\leq i \leq s+1-j$. In \cite{Calderbank}, it is also shown that a parity-check matrix for $\C'$ is of the form 
\begin{equation}\label{Hmatrix}
H=\left(\begin{array}{ccccccc}
H_{1,1}& H_{2,1} & H_{3,1} & \cdots & H_{s,1} &Id_{n-t} \\
pH_{1,2} &  pH_{2,2} & pH_{3,2} &  \cdots & pId_{t_s}  & \zero \\
\vdots & \vdots &\vdots &  \reflectbox{$\ddots$} &  \zero  & \zero \\
p^{s-2}H_{1,s-1} & p^{s-2}H_{2,s-1} & p^{s-2}Id_{t_3} &\cdots & \zero  & \zero \\
p^{s-1}H_{1,s} & p^{s-1}Id_{t_2} & \zero &\cdots& \zero  & \zero \\

\end{array}\right),
\end{equation}
where the column blocks have the same size as in (\ref{Gmatrix}).

\medskip
In this paper, we show how to construct a parity-check matrix of a $\Z_{p^s}$-additive code from a generator matrix in standard form, as a generalization of matrix (\ref{eq:HSF-4}) for quaternary linear codes. This paper is organised as follows. In Section \ref{Section:Parity}, we describe two methods to obtain a parity-check matrix, one is based on the computation of block-minors of an associated matrix to the generator matrix of the $\zps$-additive code in standard form, and another one is based on computing the block-minors in a resursive way using previous computed matrices. 
In Section \ref{section:Performance}, we describe algorithms for both methods. We also compare the computational time of an implementation of these algorithms in Magma, for some values of the parameters, together with the implementation given by a function in Magma that works for any linear code over a finite ring. Finally, in Section \ref{section:Conclusions}, we give some conclusions and future research on this topic.


\section{Computation of a parity-check matrix}\label{Section:Parity}

In this section, we present two different approaches to construct  a parity-check matrix for $\zps$-additive codes from a generator matrix in standard form (see Theorems \ref{teo: Onotation} and \ref{prop: Hnotation}). First, we present some results on the computation of determinants for square matrices with a certain structure. 


Let $A=(a_{r,s})_{1\leq r,s\leq n}$ be a square $n\times n$ matrix. It is well-known that the determinant of $A$ can be computed as 
\begin{equation}\label{eq:reg_det}
|A|=\sum_{\sigma\in \cS_{n}}\operatorname{sgn}(\sigma) \prod_{h=1}^{n}a_{h,\sigma({h})},
\end{equation}
where $\operatorname{sgn}(\sigma)$ is the parity of $\sigma$. If $A$ is a square $n\times n$ matrix of the form
\begin{equation} \label{eq:MatrixA}
\left(\begin{array}{ccccc}
a_{1,1} & a_{1,2} & \cdots & a_{1,n-1} & a_{1,n}\\
1 & a_{2,2} & \cdots & a_{2,n-1} & a_{2,n}\\
0 & 1 & \cdots & a_{3,n-1} & a_{3,n}\\
\vdots & \vdots & \ddots & \vdots & \vdots \\
0 & 0 & \cdots & 1 & a_{n,n}
\end{array}\right),
\end{equation}
then we can improve Equation (\ref{eq:reg_det}) for the determinant of $A$ computing the additions of the products that do not consider any element under the diagonal with ones. Therefore, we just need to consider permutations $\sigma\in \cS_n$ such that $\sigma(h)\geq h-1$ for all $h\in \{1,\dots,n\}$. Moreover, we can avoid multiplying by one by considering only the products of elements $a_{i,j}$ with $j\geq i$. In order to write this more formally, we introduce the following definitions.

\begin{definition}\label{defi:J}
Let $\hat{\cS}_n = \{\sigma\in \cS_{n}\,|\,\sigma(h)\geq h-1 \text{ for all } h\in \{1,\dots,n \} \}$, and
$J_\sigma=\{h_1,\dots,h_r\}=\{h\in\{1,\dots,n\} \;|\; \sigma(h)\geq h\}$. Note that $J_\sigma$ is not empty, since $\sigma(1)\geq 1$ for any $\sigma \in \hat{\cS}_n$.  We consider the elements in $J_\sigma$ ordered, i.e., $h_1<h_2<\dots < h_r$.    
\end{definition}

\begin{proposition} \label{prop:DetA}
Let $A$ be a matrix as in (\ref{eq:MatrixA}). Then, the determinant of $A$ is given by 
\begin{align*}
    |A|=&\sum_{\sigma \in \hat{\cS}_n}\operatorname{sgn}(\sigma) \prod_{h\in J_\sigma}a_{h,\sigma(h)},
\end{align*}
where $\hat{\cS}_n$ and $J_{\sigma}$ are as in Definition \ref{defi:J}.
\end{proposition}

\begin{proof}
Straightforward from (\ref{eq:reg_det}) and Definition \ref{defi:J}.
\end{proof}

\begin{example} Consider the matrix $A$ as in (\ref{eq:MatrixA}) with $n=3$, that is,
\begin{equation*} 
A=\left(\begin{array}{ccc}
a_{1,1} & a_{1,2} & a_{1,3}\\
1 & a_{2,2}  & a_{2,3}\\
0 &  1 & a_{3,3}
\end{array}\right).
\end{equation*}
In this case, we have that $\cS_3=\{Id,(1,2),(1,3),(2,3),(1,2,3),(1,3,2)\}$ and $\hat{\cS_3}=$ $\{Id,(1,2),$ $(2,3),(1,3,2)\}$. Then, $J_{Id}=\{1,2,3\}$, $J_{(1,2)}=\{1,3\}$, $J_{(2,3)}=\{1,2\}$, and $J_{(1,3,2)}=\{1\}$. Therefore,
$$
|A|=a_{1,1}a_{2,2}a_{3,3}-a_{1,2}a_{3,3}-a_{1,1}a_{2,3}+a_{1,3}.
$$
\end{example}

Now, we give an alternative expression for the computation of the determinant of a matrix A as in (\ref{eq:MatrixA}) by using the minors of the diagonal of $A$.

\begin{definition}
\label{def: minorsEscalar}
Let $A=(a_{r,s})_{1\leq r,s\leq n}$ be a square $n\times n$ matrix. The \textit{$i$-th minor of the diagonal of $A$ of order $j$}, denoted by $O_j^i$, is the determinant of the $i$-th submatrix of size $j$ in the diagonal of $A$, that is,
$$O_j^i=|(a_{r,s})_{i\leq r,s\leq i+j-1}|.$$
We consider that $O_0^i=1$ for all $i\geq 1$. Note that $O_1^i=a_{i,i}$ and $O_n^1=|A|$.
\end{definition}

\begin{proposition}\label{Prepro:unit}
Let $A$ be a matrix as in (\ref{eq:MatrixA}). Then, the determinant of $A$ is given by
\begin{equation*}
|A|=O_n^1=\sum_{k=1}^{n}(-1)^{k-1}a_{1,k}O_{n-k}^{k+1},
\end{equation*}
where $O_j^i$ is the $i$-th minor of the diagonal of $A$ of order $j$. 
\end{proposition}

\begin{proof}
Using the Laplace expansion on the first column, the determinant $|A|$ is as follows:
$$
|A|=a_{1,1}O_{n-1}^2-|A'_{n-1}|,
$$
where
$$
A'_{n-i}=\left(\begin{array}{ccccc}
a_{1,i+1} & a_{1,i+2} & \cdots & a_{1,n-1} & a_{1,n}\\
1 & a_{i+2,i+2} & \cdots & a_{i+2,n-1} & a_{i+2,n}\\
0 & 1 & \cdots & a_{i+3,n-1} & a_{i+3,n}\\
\vdots & \vdots & \ddots & \vdots & \vdots \\
0 & 0 & \cdots & 1 & a_{n,n}
\end{array}\right).
$$
We can repeat this process with $A'_{n-1}$ obtaining $|A|=a_{1,1}O_{n-1}^2-a_{1,2}O_{n-2}^3+|A'_{n-2}|$, and so on so forth until we have 
$$
|A|=\sum_{k=1}^{n}(-1)^{k-1}a_{1,k}O_{n-k}^{k+1}.
$$
\end{proof}

\begin{corollary}\label{cor:GenMinunit}
Let $A$ be a matrix as in (\ref{eq:MatrixA}). Then, 
\begin{equation}
    O_j^i= \sum_{k=i}^{i+j-1}(-1)^{i-k}a_{i,k}O_{i+j-1-k}^{k+1},
\end{equation}
where $O_j^i$ is the $i$-th minor of the diagonal of $A$ of order $j$, for all $1\leq i\leq n$ and $i\leq j\leq n$.
\end{corollary}

From now on, we consider block-matrices, that is, matrices whose entries are submatrices instead of scalars. Indeed, we first define the reduced associated matrix $G^{RA}$ of a generator matrix $G$ in standard form, which is a block-matrix. 

\begin{definition}
Let $G$ be the generator matrix of a $\Z_{p^s}$-additive code $\C$ in standard form, that is, as in (\ref{Gmatrix}). The \textit{reduced associated matrix} $G^{RA}$ of $G$ is the matrix
\begin{equation}\label{Eq:GRedAss}
G^{RA}=\left(\begin{array}{ccccc}
A_{1,2} & A_{1,3} &\cdots & A_{1,s} & A_{1,s+1} \\
Id_{t_2} & A_{2,3} & \cdots & A_{2,s} & A_{2,s+1}\\
\zero & Id_{t_3} & \cdots & A_{3,s} & A_{3,s+1}\\
\vdots & \vdots & \ddots & \vdots & \vdots\\
\zero & \zero & \cdots & Id_{t_{s}} & A_{s,s+1}
\end{array}\right).
\end{equation}
\end{definition}

In general, a blockwise determinant of a square block-matrix, computed by performing multiplications and additions of blocks, is not well-defined due to the non-compatibility of the different dimensions of each block and the fact that the product of matrices is non-commutative.  However, we propose a notion of determinant, called block-determinant, for any block-matrix of the form as in (\ref{Eq:GRedAss}), by defining an analogous expression to the one given in Proposition \ref{prop:DetA}. 
Since the block-submatrices $(A_{r,s+1})_{i\leq r,s\leq i+j-1 }$ are also in the form of (\ref{Eq:GRedAss}), we can also provide a notion of block-minors of the block-diagonal of $G^{RA}$, analogous to the minors $O_j^i$ described in Definition \ref{def: minorsEscalar}. Then, we give an analogue of Proposition \ref{Prepro:unit} to obtain another expression to compute these block-minors.

\begin{definition} \label{def:BlockDet}
Let $A$ be a block matrix as in (\ref{Eq:GRedAss}).
The $i$-th \textit{block-minor} of the block-diagonal of $A$ of order $j$, denoted also by $O_j^i$, is the block-determinant of the $i$-th submatrix of size $j$ in the block-diagonal of $A$, that is,
\begin{align*}
    O_j^i=& \left| (A_{r,s+1})_{i\leq r,s\leq i+j-1 } \right| =\left|\begin{array}{ccccc}
        A_{i,i+1} & A_{i,i+2}   & \cdots & A_{i,i+j-1} & A_{i,i+j}\\
        Id_{t_{i+1}}        & A_{i+1,i+2} & \cdots & A_{i+1,i+j-1} & A_{i+1,i+j}\\
            \zero      & Id_{t_{i+2}}          & \cdots & A_{i+2,i+j-1} & A_{i+2,i+j}\\
            \vdots     & \zero            & \ddots & \vdots        & \vdots \\
            \zero      & \dots            & \zero      &  Id_{t_{i+j-1}}           & A_{i+j-1,i+j}
    \end{array}\right|\\
\end{align*}

\begin{align}
\label{eq: detBlocks}
    =&\sum_{\sigma \in \hat{\cS}_j}\operatorname{sgn}(\sigma) \prod_{h\in J_\sigma}A_{i+h-1,i+\sigma(h)},
\end{align}
where $\hat{\cS}_j$ and $J_{\sigma}$ are as in Definition \ref{defi:J}. 
\end{definition}

The following results are used to show that the products in (\ref{eq: detBlocks}) are well-defined. 

\begin{remark} \label{remark:sigma}
    Let $\sigma\in \hat{\cS}_j$ and $h\in\{1,\dots,j\}$. Then, 
    \begin{enumerate}
    \item \label{remark:sigmaConsecutive} $\sigma(h)=h-1$ if $h\not\in J_\sigma$, 
    \item \label{remark:sigmaInverse} $\sigma^{-1}(h)\leq h+1$.
    \end{enumerate}
\end{remark}

\begin{lemma}
\label{lemma: consecutive}
Let $\sigma\in\hat{\mathcal{S}}_j$ and $J_\sigma=\{h_1,\dots,h_r\}$. Then, for any $k\in\{1,\dots,r-1\}$,  $h_{k+1}=\sigma(h_{k})+1$.
\end{lemma}

\begin{proof} 
First, we see that $\sigma(h_k)+1\in J_\sigma$. Assume $\sigma(h_k)+1\not\in J_\sigma$. By Remark \ref{remark:sigma}-\ref{remark:sigmaConsecutive}, we have that $\sigma(\sigma(h_k)+1)=(\sigma(h_k)+1)-1=\sigma(h_k)$. Therefore, since $\sigma$ is a one-to-one map, $\sigma(h_k)+1=h_k$, that is, $\sigma(h_k)=h_k-1$ which is not possible because $h_k\in J_\sigma$.  Therefore, $\sigma(h_k)+1\in J_\sigma$, and the result follows if we prove that  $\{h_{k}+1,\dots,\sigma(h_{k})\} \cap  J_\sigma =\emptyset$.

 Consider $i\in\{1,\dots,h_{k}-1\}$.  By Remark \ref{remark:sigma}-\ref{remark:sigmaInverse},    $\sigma^{-1}(i)\leq i+1 \leq h_{k}$.  If $\sigma^{-1}(i)=h_k$, then $\sigma(h_k)=i\leq h_k-1$ which is not possible since $h_k\in J_\sigma$. Therefore, $\sigma^{-1}(i)\leq h_{k}-1$. Since there are $h_{k}-1$ different values of both $\sigma^{-1}(i)$ and $i \in \{1,\dots,h_{k}-1\}$, we have that $\sigma^{-1}(h_k)\not\in\{1,\dots,h_k-1\}$. By Remark \ref{remark:sigma}-\ref{remark:sigmaInverse}, 
 $\sigma^{-1}(h_k)\leq h_k+1$. Thus, there are only two possible values remaining for $\sigma^{-1}(h_{k})$, say $h_{k}$ or $h_{k}+1$.

First, consider the case $\sigma(h_{k})=h_{k}$. Since $h_k\in J_\sigma$, we have seen that $\sigma(h_{k})+1\in J_\sigma$ and hence $h_{k}+1\in J_\sigma$. Then, since $h_k$ and $h_{k+1}$ are consecutive elements in $J_\sigma$, necessarily, $h_{k+1}=h_{k}+1=\sigma(h_k)+1$ and the result holds. Finally, consider the case $\sigma(h_{k}+1)=h_k$. In this case, $h_k+1\not\in J_\sigma.$ If $h_k+1=\sigma(h_k)$, then  clearly $\{h_{k}+1,\dots,\sigma(h_{k})\} \cap J_\sigma=\emptyset$ and we are done. If $h_k+1\not=\sigma(h_k)$, by the same argument as before, $\sigma^{-1}(h_{k}+1)\not\in\{1,\dots,h_k-1\}$. By Remark \ref{remark:sigma}-\ref{remark:sigmaInverse}, $\sigma^{-1}(h_{k}+1)\leq h_k+2$. Moreover, we have that $\sigma^{-1}(h_{k}+1)$ cannot be $h_{k}+1$ or $h_k$, so  $\sigma^{-1}(h_{k}+1)=h_{k}+2$. This implies that $h_{k}+2\not\in J_\sigma$. This argument can be applied recursively, obtaining $\{h_k+1,\dots,\sigma(h_{k})\} \cap J_\sigma =\emptyset$. 
\end{proof}

Using Lemma \ref{lemma: consecutive}, we see that all products in (\ref{eq: detBlocks}) are of the form $$A_{i+h_k-1,i+\sigma(h_k)}A_{i+\sigma(h_k),i+\sigma(h_{k+1})},$$ for any $k\in \{1,\dots,r-1\}$. Thus, $O_j^i$ is a 
well-defined matrix of size $t_i\times z$, where $z$ is the amount of columns that the matrices $A_{*,i+j}$ have, in this case $t_{i+j}$.
We consider that $O_0^i=Id$ for all $i\geq 1$. Note that $O_1^i=A_{i,i+1}$.

\begin{example}
Let $G^{RA}$ be a block matrix as in (\ref{Eq:GRedAss}). By Definition \ref{defi:J}, we have $\hat{\mathcal{S}}_2=\{Id,(1,2)\}$, $J_{Id}=\{1,2\}$ and $J_{(1,2)}=\{1\}.$ Then, the block-minor $O_2^{s-1}$ can be computed as follows:
\begin{align*}
    O_2^{s-1}=\left|\begin{array}{cc} A_{s-1,s} & A_{s-1,s+1} \\ Id_{t_s} & A_{s,s+1} \end{array}\right| 
    & =A_{s-1,s}A_{s,s+1}-A_{s-1,s+1}.
\end{align*}
Clearly,  $A_{s-1,s}A_{s,s+1}$ is a $t_{s-1}\times(n-t)$ matrix, where $n$ is the total amount of columns of $G$ and $t=\sum_{i=0}^st_i$.

In order to compute the block-minor $O_{4}^{s-3}$, we consider the set of permutations $\hat{\mathcal{S}}_4=\{Id,(3,4),(2,3),(2,4,3),(1,2),(1,2)(3,4),$ $(1,3,2), (1,4,3,2)\}$. The corresponding sets of indices given in Definition \ref{defi:J} are: $J_{Id}=\{1,2,3,4\}$, $J_{(3,4)}=\{1,2,3\}$, $J_{(2,3)}=\{1,2,4\}$, $J_{(2,4,3)}=\{1,2\}$, $J_{(1,2)}=\{1,3,4\}$, $J_{(1,2)(3,4)}=\{1,3\}$, $J_{(1,3,2)}=\{1,4\}$, and $J_{(1,4,3,2)}=\{1\}$. Then the block-minor $O_{4}^{s-3}$ is
\begin{align*}
    O_{4}^{s-3}=&\left|\begin{array}{cccc}
        A_{s-3,s-2} & A_{s-3,s-1} & A_{s-3,s} & A_{s-3,s+1} \\
         Id_{t_{s-2}} & A_{s-2,s-1} & A_{s-2,s} & A_{s-2,s+1} \\
           \zero         &  Id_{t_{s-1}} & A_{s-1,s} & A_{s-1,s+1} \\
               \zero      &   \zero          & Id_{t_s}  & A_{s,s+1}   \\
    \end{array} \right| \\
    =& A_{s-3,s-2}A_{s-2,s-1}A_{s-1,s}A_{s,s+1}-A_{s-3,s-2}A_{s-2,s-1}A_{s-1,s+1}\\
    &-A_{s-3,s-2}A_{s-2,s}A_{s,s+1}+A_{s-3,s-2}A_{s-2,s+1}-A_{s-3,s-1}A_{s-1,s}A_{s,s+1}\\&+ A_{s-3,s-1}A_{s-1,s+1}+A_{s-3,s}A_{s,s+1}-A_{s-3,s+1}.
\end{align*}
\end{example}

\begin{proposition}\label{Prepro}
Let $A$ be a block matrix as in (\ref{Eq:GRedAss}). Then, the block-determinant of $A$ is given by
\begin{equation*}
|A|=O_s^1=\sum_{k=1}^{s}(-1)^{k-1}A_{1,k+1}O_{s-k}^{k+1},
\end{equation*}
where $O_j^i$ is the $i$-th block-minor of the block-diagonal of $A$ of order $j$. 
\end{proposition}

\begin{proof}
It is easy to prove this statement by following an analogous argument to that used in the proof of Proposition \ref{Prepro:unit}.
\end{proof}

\begin{corollary}\label{cor:GenMin}
Let $A$ be a block matrix as in (\ref{Eq:GRedAss}). Then, 
\begin{equation}\label{eq: GenericOij}
    O_j^i= \sum_{k=i}^{i+j-1}(-1)^{i-k}A_{i,k+1}O_{i+j-1-k}^{k+1},
\end{equation}
where $O_j^i$ is the $i$-th block-minor of the block-diagonal of $A$ of order $j$, for all $1\leq i\leq s$ and $i\leq j\leq s$.
\end{corollary}

Now, we give the result that allows us to compute a parity-check matrix using the block-minors of the reduced associated matrix $G^{RA}$ of $G$.


\begin{theorem} \label{teo: Onotation}
Let $\C$ be a $\Z_{p^s}$-additive code of type $(n;t_1,\ldots,t_s)$ with a generator matrix $G$ as in (\ref{Gmatrix}), and $G^{RA}$ its reduced generator matrix. Then, the transpose of a parity-check matrix $H$ of $\C$ is as follows:
\begin{equation}\label{Hmatrix: minors}
H^T=\left(\begin{array}{cccccc}
H_{1,1} & pH_{1,2} & \cdots & p^{s-3}H_{1,s-2} & p^{s-2}H_{1,s-1} & p^{s-1}H_{1,s} \\
H_{2,1} & pH_{2,2} & \cdots & p^{s-3}H_{2,s-2} & p^{s-2}H_{2,s-1} & p^{s-1}Id_{t_2}\\
H_{3,1} & pH_{3,2} & \cdots & p^{s-3}H_{3,s-2} & p^{s-2}Id_{t_3} & \zero \\
H_{4,1} & pH_{4,2} & \cdots & p^{s-3}Id_{t_4} & \zero & \zero \\
\vdots & \vdots & \reflectbox{$\ddots$} & \vdots & \vdots & \vdots \\
H_{s,1} & pId_{t_s} & \zero & \zero & \zero & \zero \\
Id_{n-t} & \zero & \zero & \zero & \zero & \zero \\
\end{array}\right),
\end{equation}
where $t=\sum_{i=1}^st_i$,  
\begin{equation}\label{eq:Hijminors}
H_{i,j}=(-1)^{s+2-i-j}O^i_{s+2-i-j},
\end{equation}
for all $1\leq j \leq s$ and $1\leq i \leq s+1-j$,
and $O_k^i$ is the $i$-th block-minor of the block-diagonal of $G^{RA}$ of order $k$.  
\end{theorem}

\begin{proof}
Let $\C'$ be the $\Z_{p^s}$-additive code generated by matrix $H$ given in (\ref{Hmatrix: minors}). First, we prove that $GH^T=(\zero)$ and hence $\C'\subseteq \C^\perp$. Denote $G_s$ and $H_s$ the matrices $G$ and $H$, respectively, corresponding to the value $s$. We prove that $G_sH_s^T=(\zero)$ by induction on $s\geq 2$. For $s=2$, we have
\begin{align*}
    &H_{2,1}=-A_{2,3},\\
    &H_{1,1}=-A_{1,3}-A_{1,2}H_{2,1}=-A_{1,3}+A_{1,2}A_{2,3},\\
    &H_{1,2}=-A_{1,2}.
\end{align*}
Clearly,
$$
G_2H_2^T=\left(\begin{array}{ccc}
Id_{t_1} & A_{1,2} & A_{1,3} \\
\zero & pId_{t_2} & pA_{2,3}  
\end{array}\right)\left(\begin{array}{cc}
(A_{1,2}A_{2,3}-A_{1,3}) & -pA_{1,2} \\
-A_{2,3} & pId_{t_2}\\
Id_{n-t_1-t_2} & \zero
\end{array}\right)=
(\zero).
$$
By induction hypothesis, we assume that $G_kH_k^T=(\zero)$ for every integer $k\leq s-1$. Let us decompose the matrices $G_s$ and $H_s$ in terms of $G_{s-1}$ and $H_{s-1}$:
$$
G_s=\left(\begin{array}{cccccc}
 & & & & A_{1,s+1}\\
 & & G_{s-1} & & pA_{2,s+1}\\
 & & & & p^2A_{3,s+1}\\
 & & & & \vdots\\
\zero & \cdots & \zero & p^{s-1}Id_{t_s} & p^{s-1}A_{s,s+1}\\
\end{array}\right),
$$
$$
H_s^T=\left(\begin{array}{cccccc}
H_{1,1} & && \\
H_{2,1}& && \\
H_{3,1}& &pH_{s-1}^T&\\
H_{4,1} & && \\
\vdots & && \\
H_{s,1} & && \\
Id_{n-t} & \zero & \cdots & \zero \\
\end{array}\right).
$$
We can separate the product $G_sH_s^T$ in two parts. On the one hand, we have that
$$
G_s\left(\begin{array}{c}
pH_{s-1}^T\\
\zero\\
\end{array}\right)=\left(\begin{array}{c}
pG_{s-1}H_{s-1}^T\\
p^{s-1}Id_{t_s}\cdot pId_{t_s}\\
\end{array}\right)=(\zero)
$$
by the induction hypothesis and the structure of $H_{s-1}^T$. On the other hand,  we have to prove that
$$
G_{s}\left(\begin{array}{c}
H_{1,1}\\
H_{2,1}\\
H_{3,1}\\
H_{4,1}\\
\vdots \\
H_{s,1} \\
Id_{n-t}\\
\end{array}\right)=G_sH'=(\zero),
$$
where $H'$ is the matrix having the first $n-t$ columns of $H_s^T$.
Note that the $i$-th block-row of $G_s$ is of the form
$$
(G_s)_i=\left(\begin{array}{ccccccc}\zero &\cdots &\zero &p^{i-1}Id_{t_i} &p^{i-1}A_{i,i+1}& \cdots &p^{i-1}A_{i,s+1}\end{array}\right)
$$ 
for all $1\leq i \leq s$.
Then, 
\begin{align*}
    (G_s)_iH'& = p^{i-1}Id_{t_i}H_{i,1}+\sum_{j=1}^{s-i}p^{i-1}A_{i,i+j}H_{i+j,1}+p^{i-1}A_{i,s+1} \\
             & = p^{i-1}(-1)^{s+1-i}O_{s+1-i}^i+p^{i-1}\sum_{j=1}^{s-i}A_{i,i+j}(-1)^{s+1-i-j}O_{s+1-i-j}^{i+j}+p^{i-1}A_{i,s+1}\\
             & = p^{i-1}\left[(-1)^{s+1-i}O_{s+1-i}^i+\sum_{j=1}^{s+1-i}A_{i,i+j}(-1)^{s+1-i-j}O_{s+1-i-j}^{i+j}\right]\\
             & = p^{i-1}(-1)^{s+1-i}\left[O_{s+1-i}^i-\sum_{j=1}^{s+1-i}A_{i,i+j}(-1)^{j-1}O_{s+1-i-j}^{i+j}\right],
\end{align*}
where the second equality is given by (\ref{eq:Hijminors}) and the third one by considering that, by definition, $O_0^{s+1}=Id$. Now, 
using Corollary \ref{cor:GenMin}, it is easy to see that
$$
O_{s+1-i}^i=\sum_{j=1}^{s+1-i}A_{i,i+j}(-1)^{j-1}O_{s+1-i-j}^{i+j}.
$$
Thus $(G_s)_iH'=(\zero)$ for all $1\leq i\leq s$ and we conclude that $G_sH^T_s=(\zero)$.

We have proven that $\C'\subseteq \C^\perp$. For the other inclusion, we consider an arbitrary codeword $\mathbf{c}=(c_1,c_2,\dots,c_n)\in \C^\perp$ and we have to see that it also belongs to $\C'$. The following argument was adapted from \cite[Proposition 1.2]{Wan}. The first $n-t$ rows of $H$ have the identity matrix in the last $n-t$ columns. Therefore, we can add a linear combination of these $n-t$ first rows to $\mathbf{c}$ to obtain a new codeword of $\C^\perp$ of the form
\begin{equation*}
    \mathbf{c}^{(1)}=(c_1,c_2,\dots,c_t,0,\dots,0).
\end{equation*}
Since $\mathbf{c}^{(1)}\in \C^\perp$ is orthogonal to the last $t_s$ rows of $G_s$, we obtain
\begin{align*}
\mathbf{c}^{(1)}((G_{s})_s)^T & =\mathbf{c}^{(1)}(\zero  \ \cdots \ \zero \quad p^{s-1}Id_{t_s} \quad  p^{s-1}A_{s,s+1})^T\\
&=p^{s-1}(c_{t_1+\cdots+t_{s-1}+1},\dots,c_{t})=\zero.
\end{align*}
This means that the components $c_{t_1+\cdots+t_{s-1}+1},\dots,c_{t}$ are all multiples of $p$. The next $t_s$ rows of $H$, which are
\begin{equation}\label{eq:submatrix1}
\left(\begin{array}{cccccc}
pH_{1,2} & pH_{2,2} &  \dots & pH_{s-1,2}  & pId_{t_s} & \zero \\
\end{array}\right),
\end{equation}
have zero entries in the last $n-t$ columns and $pId_{t_s}$ in the previous $t_s$ columns. Therefore, we can add a linear combination of the rows in (\ref{eq:submatrix1}) to obtain a new codeword of $\C^\perp$ of the form
\begin{equation*}
    \mathbf{c}^{(2)}=(c_1,c_2,\dots,c_{t_1+\cdots+t_{s-1}},0,\dots,0).
\end{equation*}
Since $\mathbf{c}^{(2)}\in \C^\perp$ is orthogonal to the second to last $t_{s-1}$ rows of $G$, we obtain
$$\mathbf{c}^{(2)}((G_s)_{s-1})^T=p^{s-2}(c_{t_1+\cdots+t_{s-2}+1},\dots,c_{t_1+\cdots+t_{s-1}})=\zero.$$
This means that the components $c_{t_1+\cdots+t_{s-2}+1},\dots,c_{t_1+\cdots+t_{s-1}}$ are multiples of $p^2$. The same argument can be applied iteratively until we obtain the codeword of $\C^\perp$,
$$\mathbf{c}^{(s)}=(c_1,\dots,c_{t_1},0,\dots,0).$$
Since $\mathbf{c}^{(s)}\in \C^\perp$ is orthogonal to the first $t_1$ rows of $G$, we obtain
$$
\mathbf{c}^{(s)}((G_s)_1)^T=(c_1,\dots,c_{t_1})=\zero.
$$
Hence $\mathbf{c}^{(s)}=\zero$ and $\mathbf{c}\in \C'$, since it can be obtained by a linear combination of the rows of $H$. Consequently, $H$ is a generator and a parity-check matrix for $\C^\perp$ and $\C$, respectively.
\end{proof}

\medskip

\begin{example}\label{ex:parity1}
Let $p=2$ and $s=2$. Let $G$ be the generator matrix in standard form of a $\Z_4$-additive code $\C$ of type $(n;t_1,t_2)$ and $G^{RA}$ its reduced associated matrix:
$$
G=\left(\begin{array}{ccc}
Id_{t_1} & A_{1,2} & A_{1,3} \\
\zero & 2Id_{t_2} & 2A_{2,3}  
\end{array}\right),\quad\,\quad
G^{RA}=\left(\begin{array}{ccc}
A_{1,2} & A_{1,3} \\
Id_{t_2} & A_{2,3}  
\end{array}\right).
$$
Then, by Theorem \ref{teo: Onotation}, the transpose of a generator matrix $H$ of $\C^\perp$ can be constructed as follows:
$$
H^T=\left(\begin{array}{cc}
     A_{1,2}A_{2,3}-A_{1,3}& -2A_{1,2}  \\
     -A_{2,3}& 2Id_{t_2}\\
     Id_{n-t_1-t_2} & \mathbf{0}
\end{array}\right),
$$
since $H_{1,1}=O^1_2=A_{1,2}A_{2,3}-A_{1,3}$, $H_{2,1}=-O^2_1=-A_{2,3}$, 
and $H_{1,2}=-O^1_1=-A_{1,2}$. 
Note that this parity-check matrix $H$ of $\C$ generates the same code as the matrix  (\ref{eq:HSF-4}). Indeed, both matrices are equal if we consider $-A_{2,3}$ instead of $A_{2,3}$. Note that, in both cases, $G$ generates the same code $\C$.
\end{example}

\begin{example}\label{ex:parity2}
Let $p=2$ and $s=3$. Let $G$ be the generator matrix in standard form of a $\Z_8$-additive code $\C$ of type $(n;t_1,t_2,t_3)$ and $G^{RA}$ its reduced associated matrix:
$$
G=\left(\begin{array}{cccc}
Id_{t_1} & A_{1,2} & A_{1,3} & A_{1,4} \\
\zero & 2Id_{t_2} & 2A_{2,3} & 2A_{2,4} \\
\zero & \zero & 4Id_{t_3} & 4A_{3,4} 
\end{array}\right),\quad\,
G^{RA}=\left(\begin{array}{ccc}
 A_{1,2} & A_{1,3} & A_{1,4} \\
 Id_{t_2} & A_{2,3} & A_{2,4} \\
 \zero & Id_{t_3} & A_{3,4} 
\end{array}\right).
$$
Then, the transpose of a generator matrix $H$ of $\C^\perp$ can be constructed as follows:
\vspace{.5cm}

\noindent
\resizebox{\linewidth}{!}{%
$
H^T=\left(\begin{array}{ccc}
-(A_{1,2}A_{2,3}A_{3,4}+A_{1,4}-A_{1,2}A_{2,4}-A_{1,3}A_{3,4}) & 2\left(A_{1,2}A_{2,3}-A_{1,3}\right) & -4A_{1,2} \\
 A_{2,3}A_{3,4}-A_{2,4} & -2A_{2,3} & 4Id_{t_2} \\
-A_{3,4} & 2Id_{t_3} & \zero \\
Id_{n-t_1-t_2-t_3} & \zero & \zero 
\end{array}\right)
$}
by Theorem \ref{teo: Onotation} and Corollary \ref{cor:GenMin}. For example, we have that $H_{1,1}=-O^1_3=A_{1,2}O^2_2-A_{1,3}O^3_1+A_{1,4}O_0^4=A_{1,2}( A_{2,3}A_{3,4} -A_{2,4})-A_{1,3}A_{3,4}+A_{1,4}$.
\end{example}

The computation of a parity-check matrix by using Theorem \ref{teo: Onotation} requires the reckoning of many minors. The computation of these minors, $O^i_{s+2-i-j}$, is carried out using Corollary \ref{cor:GenMin}, so it requires the computation of $j$ different minors of lower order. In this case, we assume that we compute each one of the blocks of the parity-check matrix $H_{i,j}$ independently. 
However, now, we show that following an appropriate order in the computation of the different matrices $H_{i,j}$, we are able to obtain an expression to compute $O_{s+2-i-j}$, where all the minors of lower order in (\ref{eq: GenericOij}) have already been computed in a previous step. In fact, we can obtain a similar expression, which directly relates $H_{i,j}$ with all others $H_{i,k}$ such that $k\geq j$. This is shown in Theorem \ref{prop: Hnotation}.


\begin{theorem}
\label{prop: Hnotation}
Let $\C$ be a $\Z_{p^s}$-additive code of type $(n;t_1,\ldots,t_s)$ with a generator matrix $G$ as in (\ref{Gmatrix}). Then, the transpose of a parity-check matrix $H$ of $\C$ is as follows:
\begin{equation}\label{Hmatrix:iterations}
H^T=\left(\begin{array}{ccccc}
H_{1,1}& pH_{1,2} & \cdots & p^{s-2}H_{1,s-1} & p^{s-1}H_{1,s} \\
H_{2,1} & pH_{2,2} & \cdots & p^{s-2}H_{2,s-1} & p^{s-1}Id_{t_2} \\
H_{3,1} & pH_{3,2} & \cdots & p^{s-2}Id_{t_3} & \zero \\
\vdots & \vdots & \reflectbox{$\ddots$} & \vdots & \vdots \\
H_{s,1} & pId_{t_s} & \zero & \zero & \zero \\
Id_{n-t} & \zero & \zero & \zero & \zero \\
\end{array}\right),
\end{equation}
where $t=\sum_{i=1}^s t_i$ and
\begin{equation}\label{eq: HijRec}
H_{i,j}=-\left(A_{i,s-j+2}+\sum_{k=i+1}^{s-j+1}A_{i,k}H_{k,j}\right)
\end{equation}
for all $1\leq j \leq s$ and $1\leq i<s-j+1$. Note that $H_{s-j+1,j}=-A_{s-j+1,s-j+2}$.
\end{theorem}

\begin{proof}
We prove this statement by seeing that the matrix given in (\ref{Hmatrix:iterations}) is the same as the one in (\ref{Hmatrix: minors}). To achieve that, we show that $H_{i,j}=\hat{H}_{i,j}$ for all $1\leq j \leq s$ and $1\leq i\leq s-j+1$, where $\hat{H}_{i,j}$ is as in (\ref{eq:Hijminors}), that is,
\begin{equation}\label{eq:induction}
H_{i,j}=-\left(A_{i,s-j+2}+\sum_{k=i+1}^{s-j+1}A_{i,k}H_{k,j}\right)=(-1)^{s+2-i-j}O^i_{s+2-i-j}=\hat{H}_{i,j}.
\end{equation}
We prove this by induction on $i$ for any $j\in\{1,\dots,s\}$. For the case $i=s+1-j$, we have that
$$
H_{s+1-j,j}=-\left(A_{s+1-j,s-j+2}\right)=(-1)O^{s+1-j}_{1}=\hat{H}_{s+1-j,j}.
$$
By induction hypothesis, we assume that (\ref{eq:induction}) is true for $i\leq s-j+1$ and we want to see that it is true for $i-1$, i.e., $H_{i-1,j}=\hat{H}_{i-1,j}$. We have that
\begin{align*}
    \hat{H}_{i-1,j}&=(-1)^{s+2-j-(i-1)}O_{s+2-j-(i-1)}^{i-1}\\
                   &=(-1)^{s+2-j-(i-1)}\sum_{k=i-1}^{s+1-j}A_{i-1,k+1}(-1)^{i-1-k}O_{s+1-j-k}^{k+1}\\
                   &=(-1)^{s+2-j-(i-1)}\sum_{k=i}^{s+2-j}A_{i-1,k}(-1)^{i-k}O_{s+2-j-k}^{k}\\
                   &=-\sum_{k=i}^{s+2-j}A_{i-1,k}(-1)^{s+2-j-k}O_{s+2-j-k}^{k}\\
                   &=-\left(A_{i-1,s+2-j}+\sum_{k=i}^{s+1-j}A_{i-1,k}(-1)^{s+2-j-k}O_{s+2-j-k}^{k}\right)\\
                   &=-\left(A_{i-1,s+2-j}+\sum_{k=i}^{s+1-j}A_{i-1,k}H_{k,j}\right)\\
                   &=H_{i-1,j}.
\end{align*}
The first equality is by definition, the second is by Corollary \ref{cor:GenMin}, the third is a rearrangement of the indices, the sixth is by the induction hypothesis, and the last one is by definition.  
\end{proof}

\section{Performance comparison}\label{section:Performance}

In this section, we describe two algorithms that implement the computation of a parity-check matrix for $\zps$-additive codes, from a generator matrix in standard form. They are based on Theorems \ref{teo: Onotation} and Theorem \ref{prop: Hnotation}, respectively. 
First, we show a naive implementation, which is based on computing each submatrix $H_{i,j}$ in (\ref{Hmatrix: minors}) by using the expression given in (\ref{eq:Hijminors}) and Corollary \ref{cor:GenMin}. Afterwards, we present an iterative construction that reduces the calculations considerably by using the expression given in (\ref{eq: HijRec}). Then, the performance of these algorithms implemented in Magma is compared with the performance if we use the current available function in Magma for codes over finite rings in general. A time computation and time complexity analysis are also given. 

\subsection{Algorithms description}

The first procedure corresponds to the one presented in Theorem \ref{teo: Onotation}, considering that each one of the blocks $H_{i,j}$ in (\ref{Hmatrix: minors}) is computed independently, 
 by using the expression given in (\ref{eq:Hijminors}), that is, $H_{i,j}=(-1)^{s+2-i-j}O^i_{s+2-i-j}$, and Corollary \ref{cor:GenMin} to compute  each one of the block-minors $O^i_{s+2-i-j}$ from the computation of different minors of lower order. 
This implementation is shown in Algorithm \ref{alg: minors}. 

\begin{algorithm}[ht]
	\caption{Parity-check matrix in standard form. Minors construction.}
	\label{alg: minors}
	\begin{algorithmic}[1]
		\REQUIRE  A $\zps$-additive code $\C$ of type $(n; t_1,\dots,t_s)$.
		\STATE Compute a generator matrix $G$ in standard form  of $\C$.
        \STATE Define $t:=t_1+\dots+t_s$.
        \STATE Define a zero matrix $H^T$ with $n$ rows and $n-t$ columns.
        \STATE Define $numCol := 1$.
		\FOR{$j := 1,\dots,s$}
            \STATE Define $numRow := 1$.
            \FOR{$i := 1,\dots, s-j+1$}
                \STATE Compute $O^i_{s+2-i-j}$ using the block matrix $G$ and Corollary \ref{cor:GenMin}.
                \STATE Define $H_{i,j}:=(-1)^{s+2-i-j}O^i_{s+2-i-j}$.
                \STATE Insert $p^{j-1}H_{i,j}$ at position $(numRow, numCol)$ in $H^T$.
                \STATE $numRow := numRow+t_i$.
            \ENDFOR
            \IF{$j=1$}
                \STATE Insert $Id_{n-t}$ at position $(numRow, numCol)$ in $H^T$.
                \STATE $numCol := numCol+n-t$.
            \ELSE
                \STATE Insert $p^{j-1}Id_{t_{s-j+2}}$ at position $(numRow, numCol)$ in $H^T$.
                \STATE $numCol := numCol+t_{s-j+2}$.
            \ENDIF
        \ENDFOR
		\RETURN  The parity-check matrix $H$.
	\end{algorithmic}
\end{algorithm} 



With the result given by Theorem \ref{prop: Hnotation}, we can easily define a new implementation, which reduces the number of operations compared to Algorithm \ref{alg: minors}. In particular, for each block-column $j$, we can compute each $H_{i,j}$ starting from $H_{s-j+1,j}= -A_{s-j+1,s-j+2}$ and using (\ref{eq: HijRec}) to obtain $H_{i,j}$ for $1\leq i< s-j+1$, in decreasing order. Since all $H_{k,j}$, for $k\geq i$,  have been already determined when $H_{i,j}$ is computed, no additional operations are performed apart from the sums and products of matrices represented in (\ref{eq: HijRec}). This new implementation is shown in Algorithm \ref{alg: iterative}.

\begin{algorithm}[ht]
	\caption{Parity-check matrix in standard form. Iterative construction.}
	\label{alg: iterative}
	\begin{algorithmic}[1]
		\REQUIRE  A $\zps$-additive code $\C$ of type $(n; t_1,\dots,t_s)$.
		\STATE Compute a generator matrix $G$ in standard form of $\C$.
        \STATE Define $t:=t_1+\dots+t_s$.
        \STATE Define a zero matrix $H^T$ with $n$ rows and $n-t$ columns.
        \STATE Define $numCol := 1$
		\FOR{$j := 1,\dots,s$}
                \STATE Define $numRow:=t_1+\dots+t_{s-j+1}+1$.
                \IF{$j=1$}
                    \STATE Insert $Id_{n-t}$ at position $(numRow, numCol)$ in $H^T$.
                \ELSE
                    \STATE Insert $p^{j-1}Id_{t_{s-j+2}}$ at position $(numRow, numCol)$ in $H^T$.
                \ENDIF
                \STATE $numRow := numRow-t_{s-j+1}$
                \STATE Define $H_{s-j+1,j}:=-A_{s-j+1,s-j+2}$.
                \STATE Insert $p^{j-1}H_{s-j+1,j}$ at position $(numRow, numCol)$ in $H^T$.
            \FOR{$i := s-j,\dots, 1$ by $-1$}
                \STATE $numRow := numRow-t_i$.
                \STATE Compute $H_{i, j}$ using (\ref{eq: HijRec}) and previously computed matrices $H_{k,j}$, for $k:=i+1,\dots,s-j+1$.
                \STATE Insert $p^{j-1}H_{i,j}$ at position $(numRow, numCol)$ in $H^T$.
            \ENDFOR
            \IF{$j=1$}
                \STATE $numCol := numCol+ n-t$.
            \ELSE
                \STATE $numCol := numCol + t_{s-j+2}$.
            \ENDIF
        \ENDFOR
		\RETURN  The parity-check matrix $H$.
	\end{algorithmic}
\end{algorithm} 


\subsection{Performance comparison}


In this subsection, we compare three different methods for computing the parity-check matrix of a $\zps$-additive code. Two of them are the different versions of the method that can be obtained from Theorem \ref{teo: Onotation} and Theorem \ref{prop: Hnotation}, which are described in Algorithm \ref{alg: minors} and Algorithm \ref{alg: iterative}, respectively. The third method consists on using the Magma function \texttt{ParityCheckMatrix()}, included in the current official distribution \cite{Magma}, which computes a parity-check matrix for a linear code defined over any finite ring. First, we make an experimental comparison by using Magma and present the results through some graphs, thereafter we calculate the complexity of the methods introduced in this paper.

\subsubsection{Computation time analysis}

In order to compare the performance of Algorithms \ref{alg: minors} and \ref{alg: iterative} and the Magma function, we consider a random $\zps$-additive code $\C$ of type $(n;\ell,\dots,\ell)$, that is, with $t_i=\ell$ for any $1\leq i \leq s$. We study the effect of changing the parameters $n$, $s$, and $\ell$ on the three different methods. The first method (Algorithm \ref{alg: minors}), which computes each minor independently, is labeled as \textit{Minors}. The second method (Algorithm \ref{alg: iterative}), which computes the minors iteratively, is labeled as \textit{Iterative}. Finally, the third method, which uses the Magma function \texttt{ParityCheckMatrix()}, is labeled as \textit{Generic}.

Figure 1 shows the computation times of all methods for random $\Z_{3^s}$-additive codes of type $(1000; 2,\dots,2)$, 
where $s$ takes values between 2 and 16.
Similarly, Figure 2 shows the computation times of all methods for random $\Z_{3^4}$-additive codes of type $(1000; \ell,\dots,\ell)$, 
where $\ell$ takes values between $2$ and $20$. Finally, Figures 3 and 4 show the computation times of all methods for random $\Z_{3^{10}}$-additive codes of type $(n; 2,\dots,2)$, 
where $n \in \{2^i\cdot 100\, \mid \, 0\leq i\leq 8\}$.


    

\begin{figure}
    \centering
    \includegraphics[width=\textwidth]{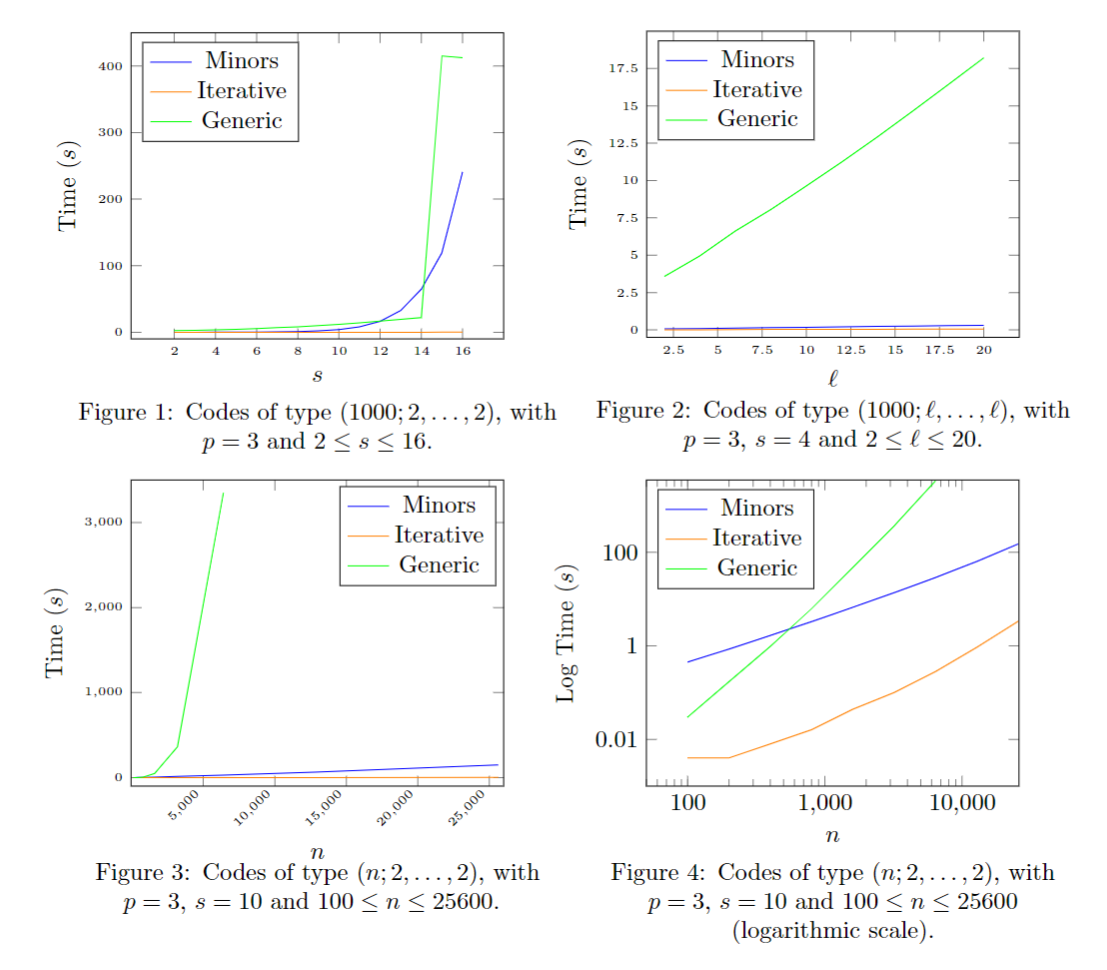}
\end{figure}

\subsubsection{Time complexity analysis}
Let us consider Algorithms \ref{alg: minors} and \ref{alg: iterative}, which are based on the following expressions, respectively:
\begin{equation}
\label{eq: minors}
    \hat{H}_{i,j}=(-1)^{s+2-i-j}O^i_{s+2-i-j}\quad\text{ and}
\end{equation}
\begin{equation}
\label{eq: recursive}
    H_{i,j}=-\left(A_{i,s-j+2}+\sum_{k=i+1}^{s-j+1}A_{i,k}H_{k,j}\right),
\end{equation}
for $1\leq j \leq s$ and $1\leq i\leq s-j+1$. Note that (\ref{eq: minors}) and (\ref{eq: recursive}) coincide with the equations given in Theorem \ref{teo: Onotation} and Theorem \ref{prop: Hnotation}, respectively. In the first case, we denote the submatrices as $ \hat{H}_{i,j}$ instead of $H_{i,j}$ in order to distinguish between both methods.

For simplicity, let us assume that $\mathcal{C}$ is a $\zps$-additive code of type $(n; \ell,\dots, \ell)$. Then, $t=s\ell$, $H_{i,1}$ is a $\ell\times (n-t)$ matrix for any $1\leq i\leq s$, and $H_{i,j}$ is a $\ell\times \ell$ matrix for any $2\leq j\leq s$ and $1\leq i\leq s-j+1$. We denote by $\hat{T}_{i,j}(s,n,\ell)$ and $T_{i,j}(s,n,\ell)$ the runtime needed to compute $\hat{H}_{i,j}$ and $H_{i,j}$, respectively. We also denote by $S(a,b)$ the computation time of the addition of two $a\times b$ matrices over $\Z_{p^s}$ and $P(a,b,c)$ the computation time of the product of an $a\times b$ matrix by a $b\times c$ matrix. 

With the aim of computing $\hat{H}_{i,j}$, we first estimate the complexity of determining any block-minor $O_j^i$. Due to the structure of $G^{RA}$, by Corollary \ref{cor:GenMin}, we can compute $O_j^i$ by calculating block-determinants of one dimension less and the same structure. Then, by induction, it easy to show that the runtime needed to compute $O_j^i$ is $(2^{s-i}-1)\left(P(\ell,\ell,n-t)+S(\ell,n-t)\right)$ for $j=s+1-i$ and $(2^{j-1}-1)\left(P(\ell,\ell,\ell)+S(\ell,\ell)\right)$ for $j<s+1-i.$ Thus,
by using (\ref{eq: minors}), we have that
\begin{align}
    &\hat{T}_{i,1}(s,n,\ell)=(2^{s-i}-1)\left(P(\ell,\ell,n-t)+S(\ell,n-t)\right),\nonumber \\
    &\hat{T}_{i,j}(s,n,\ell)=(2^{s+1-i-j}-1)\left(P(\ell,\ell,\ell)+S(\ell,\ell)\right) \text{ for } j>1. \label{newRefTtilda}
\end{align}
In order to obtain $H_{i,j}$, we need to compute $H_{i',j}$ for all $i\leq i'\leq s-j+1$. Thus, for each $1\leq j\leq s$, we start with $H_{s-j+1,j}=-A_{s-j+1,s-j+2}$ and then compute the sequence of matrices $H_{s-j,j},H_{s-j-1,j},\dots,H_{1,j}$ by using (\ref{eq: recursive}). In this case, we have that
\begin{align} 
    &T_{i,1}(s,n,\ell)=(s-i)P(\ell,\ell,n-t)+(S(\ell,n-t)),\nonumber \\
    &T_{i,j}(s,n,\ell)=(s-j-i+1)(P(\ell,\ell,\ell)+S(\ell,\ell)) \text{ for } j>1.\label{newRefT}
\end{align}
Therefore, the total runtime of computing the parity-check matrix of $\mathcal{C}$ by using Algorithms \ref{alg: minors} and \ref{alg: iterative} is given by the following results:
%

\begin{proposition}
Let $\C$ be a $\zps$-additive code of type $(n; \ell,\dots,\ell)$, and $t=s\ell$.  The total runtime of computing the parity-check matrix of $\C$ by using Algorithm \ref{alg: minors} is 
\begin{align*}
    \hat{T}(s,n,\ell)
    &=\left(2^{s}-1-s\right)\left(P(\ell,\ell,n-t)+S(\ell,n-t)\right)+\\
    &+\left(2^s-1-\frac{s^2}{2}-\frac{s}{2}\right)\left(P(\ell,\ell,\ell)+S(\ell,\ell)\right).
\end{align*}
\end{proposition}

\begin{proof}
We have  
    $\hat{T}(s,n,\ell)=\sum_{i=1}^{s}\hat{T}_{i,1}(s,n,\ell)+\sum_{j=2}^s\sum_{i=1}^{s-j+1}\hat{T}_{i,j}(s,n,\ell)$. Recall that $\sum_{k=0}^{n} 2^k=2^{n+1}-1$.
    By using (\ref{newRefTtilda}), since $\sum_{i=1}^{s}(2^{s-i}-1) =2^s-1-s$ and $\sum_{j=2}^s\sum_{i=1}^{s-j+1}(2^{s+1-i-j}-1)=\sum_{j=2}^s (2^{s+1-j}-s+j-2)=2^s-1-\frac{s^2}{2}-\frac{s}{2}$, the result follows. 
\end{proof}

\begin{proposition}
Let $\C$ be a $\zps$-additive code of type $(n; \ell,\dots,\ell)$, and $t=s\ell$.  The total runime of computing the parity-check matrix of $\C$ by using Algorithm \ref{alg: iterative} is 
    \begin{align*}
T(s,n,\ell)
    &=\frac{s(s-1)}{2}(P(\ell,\ell,n-t)+S(\ell,n-t))\\
    &+\frac{1}{6}(s^3-3s^2+2s)(P(\ell,\ell,\ell)+S(\ell,\ell)).
\end{align*}
\end{proposition}

\begin{proof}
We have  
$T(s,n,\ell)=\sum_{i=1}^{s}T_{i,1}(s,n,\ell)+\sum_{j=2}^s\sum_{i=1}^{s-j+1}T_{i,j}(s,n,\ell)$. We also have that $\sum_{i=1}^{s}(s-i)=\sum_{j=0}^{s-1}j=s(s-1)/2$ and
\begin{align*}
\sum_{j=2}^s &\sum_{i=1}^{s-j+1}(s-j-i+1)=\sum_{j=2}^s (s-j)(s-j+1)/2\\
=&\frac{1}{2}(\sum_{j=2}^s s^2 +\sum_{j=2}^s j^2 -\sum_{j=2}^s j(2s+1))\\
=&\frac{1}{2}(s^3-s + (2s^3+3s^2+s-6)/6 -(2s^3+3s^2-3s-2)/2\\
=&\frac{1}{6}(s^3-3s^2+2s).
\end{align*}
Finally, by  (\ref{newRefT}), the result follows. 
\end{proof}

\medskip
Regarding the asymptotic complexity of the algorithms, since $S(a,b)$ is $O(ab)$ and $P(a,b,c)$ is $O(abc)$, we obtain $S(\ell,n-t)+P(\ell,\ell,n-t)=O((n-t)\ell^2)$ and $S(\ell,\ell)+P(\ell,\ell,\ell)=O(\ell^3)$. Therefore,
\begin{align*}
    \hat{T}(s,n,\ell)=O(2^{s}\ell^2\left(n+s\ell \right))\\
    T(s,n,\ell)=O(s^2\ell^2n).
\end{align*}

If we only consider the variable $n$, we obtain that the algorithms are $O(n)$. Otherwise, if we take into account the variable $s$, we can see that while the first algorithm is exponential, the second proposal has square polynomial complexity, which adjust with the data shown in Figures 1, 2, 3 and 4.

\section{Conclusion}\label{section:Conclusions}

Two different methods to compute a parity-check matrix for $\zps$-additive codes have been introduced. Even though they are very similar methods, and  their performance are comparable under some conditions, we have showed that they perform very different when the parameters of the code changes. We have also established experimentally that both are better than the current algorithm included in Magma for any linear code over a finite ring.

A Magma function to compute the dual  of $\zps$-linear codes has been included in a new Magma package to deal with linear codes over $\zps$ \cite{MagmaZps}. This function is based on the construction of a parity check matrix using Algorithm \ref{alg: iterative}, and it is more efficient than the current available function in Magma for codes over finite rings in general. This new package also allows the construction of some families of $\zps$-linear codes, and includes functions related to generalized Gray maps, information sets, the process of encoding and decoding using permutation decoding, among others.  Indeed, this package generalizes some of the functions for codes over $\Z_4$, which are already included in the standard Magma distribution \cite{Magma}. It has been developed mainly by the authors of this paper and the collaboration of some undergraduate students. The first version of this new package and a manual describing all functions will be released this year, and it will be available in a GitHub repository and in the CCSG web site (\url{http://ccsg.uab.cat})

Further research is possible in different directions. A natural generalisation  would be to adapt these algorithms to compute a parity-check matrix for codes over mixed alphabets like $\mathbb{Z}_p\zps$-additive codes or even the more generic $\mathbb{Z}_p\mathbb{Z}_{p^2}\dots\zps$-additive codes. 

The generator matrix in standard form as in (\ref{Gmatrix}) has a very similar structure to the partial generator matrix  of convolutional codes (also called expanded partial generator matrix). Due to this similarity,   the methods presented in this paper are adaptable to compute  a  parity-check matrix for these codes  as long as  it exists.

\end{document}